\documentclass[5p,twocolumn,times,nopreprintline]{elsarticle}
\usepackage[numbers]{natbib}
\usepackage{amsmath}
\usepackage{booktabs}
\usepackage[colorlinks=true,pdfstartview=FitH,linkcolor=blue,anchorcolor=violet,citecolor=magenta]
{hyperref}

\usepackage{orcidlink}
\usepackage[utf8]{inputenc}
\usepackage[T1]{fontenc}
\usepackage{CJKutf8}

\newcommand{\songti}[1]{\begin{CJK*}{UTF8}{gbsn}#1\end{CJK*}}
\newcommand{\kaiti}[1]{\begin{CJK*}{UTF8}{gkai}#1\end{CJK*}}

\def\vp{\mathbf{p}}
\def\vs{\mathbf{s}}
\def\QJD{\textsf{QJD}~}
\def\RXD{\textsf{RXD}~}
\newcommand{\bookt}[1]{\textit{#1}}      
\newcommand{\cbookt}[1]{\kaiti{#1}}

\newcommand{\gudu}{\textit{gudu}}

\graphicspath{{./}}

\title{Determining the observation epochs of star catalogs from ancient China using the Generalized Hough Transform method\tnoteref{t1}}
\tnotetext[t1]{Publish in \textbf{Astronomical Techniques and Instruments} volume 2, 2024. DOI:~\href{https://doi.org/10.61977/ati2024013}{10.61977/ati2024013}}

\author[1,2]{Boliang He (\kaiti{何勃亮}~\orcidlink{0000-0002-3244-7312})}
\ead{hebl@nao.cas.cn}

\author[1,2]{Yongheng Zhao \corref{cor1} (\kaiti{赵永恒}~\orcidlink{0000-0001-5298-2833})} 
\ead{yzhao@nao.cas.cn}

\cortext[cor1]{Corresponding author}

\address[1]{National Astronomical Observatories, Chinese Academy of Sciences, 100101, China}
\address[2]{University of Chinese Academy of Sciences, Beijing, 100049, China}

\begin{document}

\begin{abstract}
    Ancient China recorded a wealth of astronomical observations, notably distinguished by the inclusion of empirical measurements of stellar observations. However, determining the precise observational epochs for these datasets poses a formidable challenge. This study employs the Generalized Hough Transform methodology to analyze two distinct sets of observational data originating from the Song and Yuan dynasties, allowing accurate estimation of the epochs of these stellar observations. This research introduces a novel and systematic approach, offering a scholarly perspective for the analysis of additional datasets within the domain of ancient astronomical catalogs in future investigations.
\end{abstract}

\begin{keyword}
    History of astronomy \sep Catalogs \sep Period determination
\end{keyword}

\maketitle

\section{INTRODUCTION}

  Records from ancient China include a wealth of astronomical observations, prominently featuring a valuable compilation of meticulously measured stellar observations. Contemporary scholars have analyzed these data extensively, but the observational epochs remain elusive. Conventional methodologies use the precession correction approach, regressing observational data with precession data~\citep{PanNai2009HCSO}. Fourier analysis methods have also been applied to estimate epochs by examining stellar data to scrutinize declination errors~\citep{sunxc1996twhc}.
  
  Previous methodologies for epoch determination required highly accurate correspondences between ancient and contemporary stars, with their precise positional coordinates. In this study, we propose the implementation of a novel algorithm using the Generalized Hough Transform to enhance the precision of calculations~\citep{hough1962method}. Unlike its predecessors, this technique operates through probabilistic calculations, alleviating the requirement for absolute accuracy in the data, with substantial correctness in the majority of the data deemed sufficient. In the computational process, only the departure from the pole degree data of the stars is essential, differentiating it from other methods that mandate not only the departure from \textit{qu ji du} \songti{去极度} (distance to the pole) but also \textit{ru xiu du} \songti{入宿度} (distance to the west boundary of \textit{xiu}).
  
  \begin{figure}[ht]
    \centering
    \includegraphics[width=0.66\columnwidth]{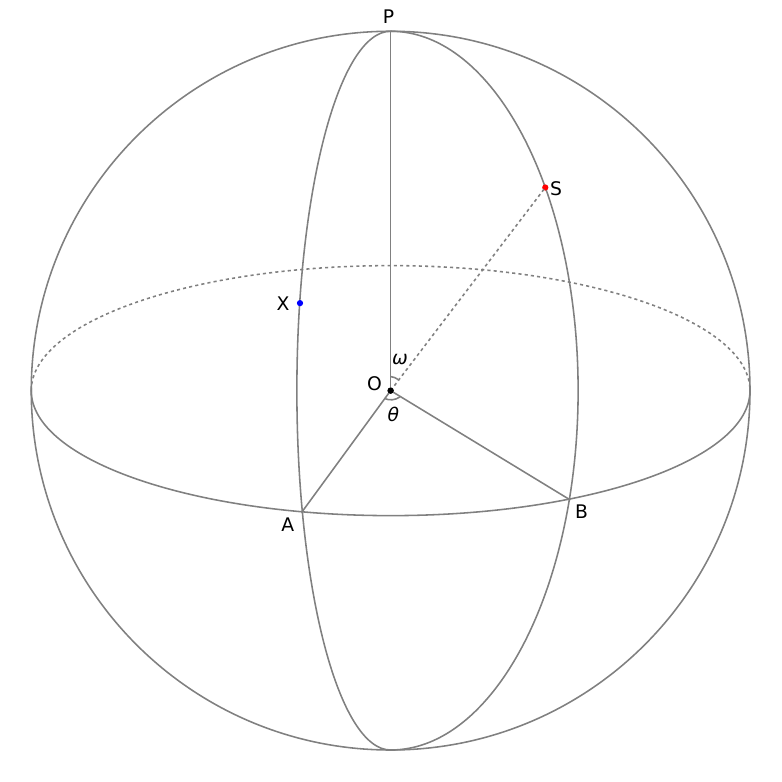}
    \caption{Illustration of an ancient Chinese equatorial coordinatesystem.}
    \label{fig:qujidu}
  \end{figure}

  This study employs the Generalized Hough Transform to analyze ancient stellar observations, aiming to precisely estimate their associated epochs.

\section{TECHNICAL METHODS}

The Hough Transform is an ingenious method that transforms a global curve detection problem into an efficient peak detection problem in parameter space. It accomplishes this by identifying imperfect instances of objects with a specific type of shape through a voting process. This voting process occurs in a parameter space, wherein candidate objects are regarded as local maxima within the accumulator space explicitly constructed by the algorithm responsible for computing the Hough Transform.

\subsection{Algorithm Principle}

The records of celestial body positions in ancient Chinese astronomy are commonly expressed in terms of two parameters. The \textit{ru xiu du} \songti{入宿度} is the distance to the west boundary of \textit{xiu}, astronomically equivalent to the right ascension of a star in an equatorial coordinate system. The only distinction between this and the modern system lies in its reference point, which is not the spring equinox but the determinative star (\textit{ju xing}) of the \textit{xiu}. This coordinate is always expressed as the number of degrees within a specific \textit{xiu}. The other parameter, the \textit{qu ji du} \songti{去极度} or distance to the pole, is equivalent to declination. Using Fig. \ref{fig:qujidu} as an illustrative example, point \textbf{X} denotes the position of a star within a specific mansion, and \textbf{P} designates the north celestial pole. Consequently, \textbf{PX} signifies the reference meridian for this mansion.

For a star S in this mansion, the difference in longitude $\theta$ represents the entry degree, while the angle $\omega$ from the north celestial pole to point S is the departure degree. \textit{Ru xiu du} (\RXD) is denoted by $\theta$, and \textit{qu ji du} (\QJD) is represented by $\omega$. The relationship between the \QJD $\omega$ and the declination ($\delta$) in the modern equatorial coordinate system is given by

$$\delta = 90^{\circ} - \omega$$

In contrast, in modern observational astronomy, stellar positions in the celestial sphere are usually articulated in terms of right ascension, $\alpha$, and declination, $\delta$, as

$$\vs = (\alpha_s, \delta_s)$$

In the \QJD data, $\omega$ signifies the angular distance of a star from the north celestial pole. The position of the north celestial pole at the observational epoch in the modern equatorial coordinate system is denoted as p = $(\alpha,\delta)$. Therefore, for any given star, the dot product of the two vectors gives

\begin{equation}
  \begin{aligned}
  \vs \cdot \vp &= \lvert\vs\rvert\lvert\vp\rvert \cos \omega \\
   &=  \cos \omega 
  \end{aligned}
\end{equation}

\begin{figure}[htbp]
    \centering
    \includegraphics[width=\columnwidth]{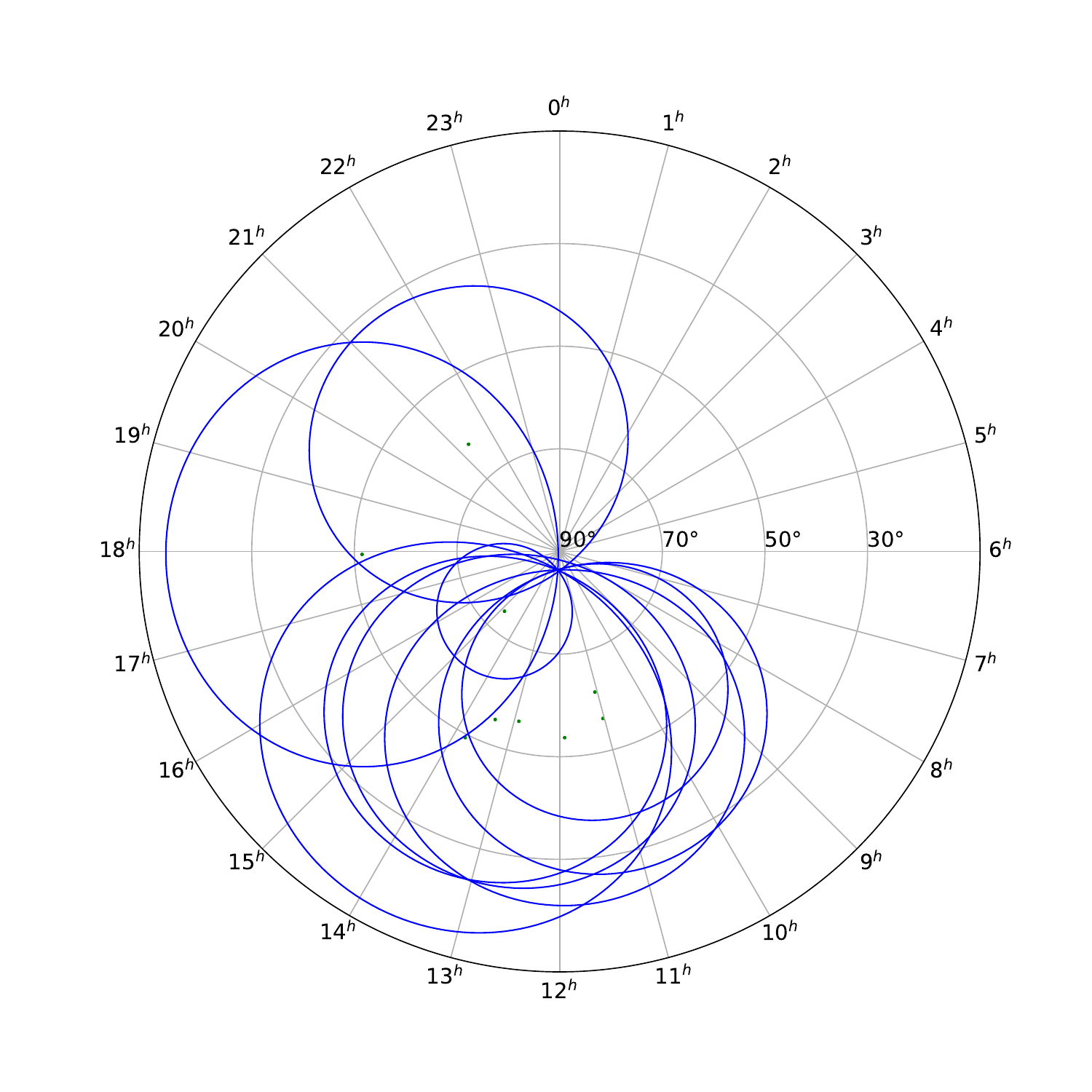}
    \caption{The parameter space of stellar positions after Hough
    Transform, showing spatial location. The blue lines depict the
    representation of nine stars in the parameter space.}\label{fig:htps}
  
  \end{figure}

In the $(\alpha,\delta)$ parameter space, the north celestial pole for each set of data corresponds to a circle, as shown in Fig. \ref{fig:htps}. Nine sets of stellar observation data near the north celestial pole have been selected, and it is evident that the majority of the circles in the parameter space converge to a singular point. Only one circle deviates slightly from the convergence point. With an increased dataset, the optimal convergence point for data can be ascertained probabilistically, serving as the candidate location for the north pole.

\begin{figure}[htbp]
    \centering
    \includegraphics[width=\columnwidth]{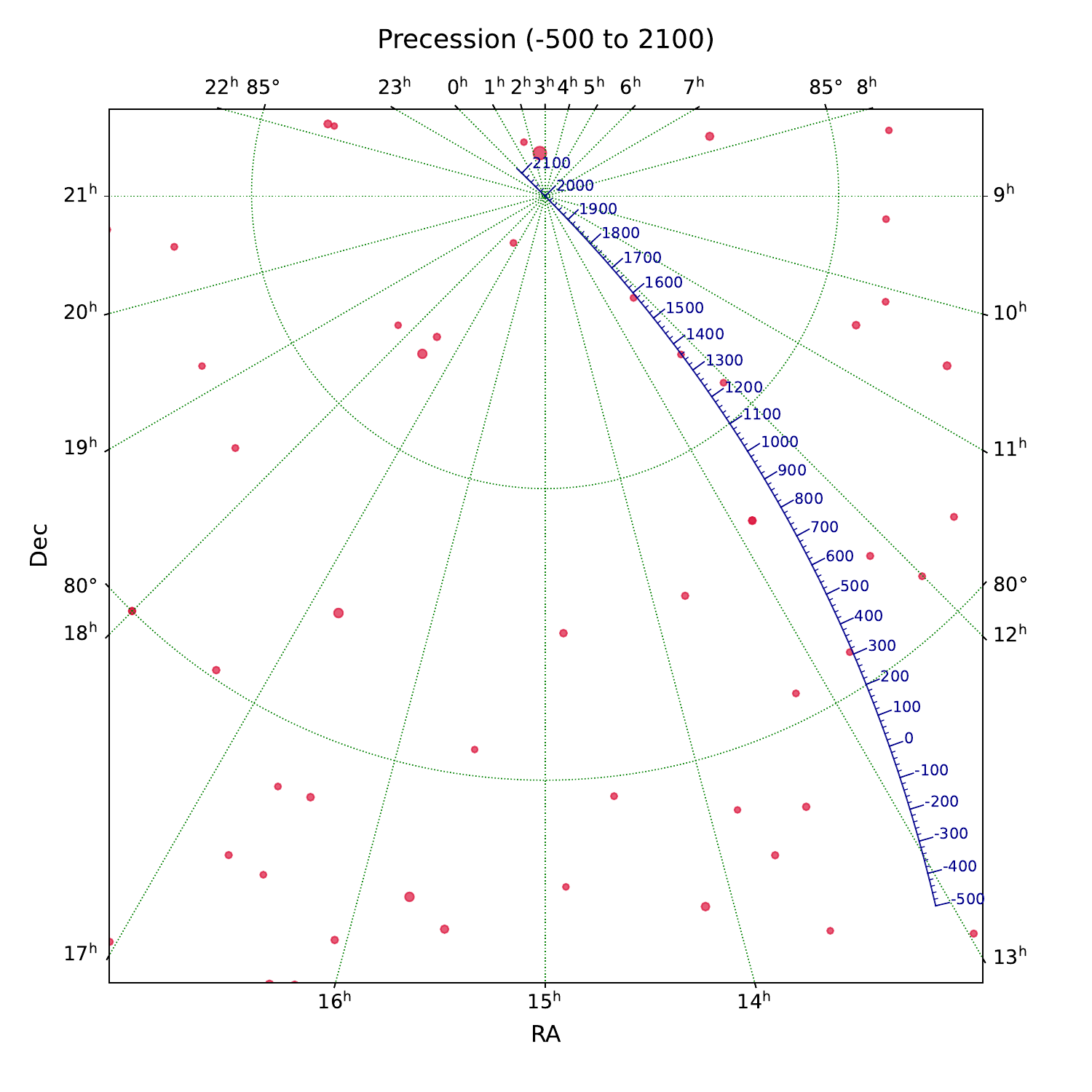}
    \caption{The position of the north celestial pole from 500 BCE to 2100 CE.}
    \label{fig:precession}
  \end{figure}

The Hough Transform can be used to process and calculate all of the stellar observation data, allowing the determination of the coordinates of the north celestial pole at the observational epoch in the modern reference frame. Subsequently, by incorporating the position of the north celestial pole along with precession data (refer to Fig. \ref{fig:precession}), the true epochs of the stellar observations can be ascertained.

\begin{figure}[htbp]
  \centering
  \includegraphics[width=0.9\columnwidth]{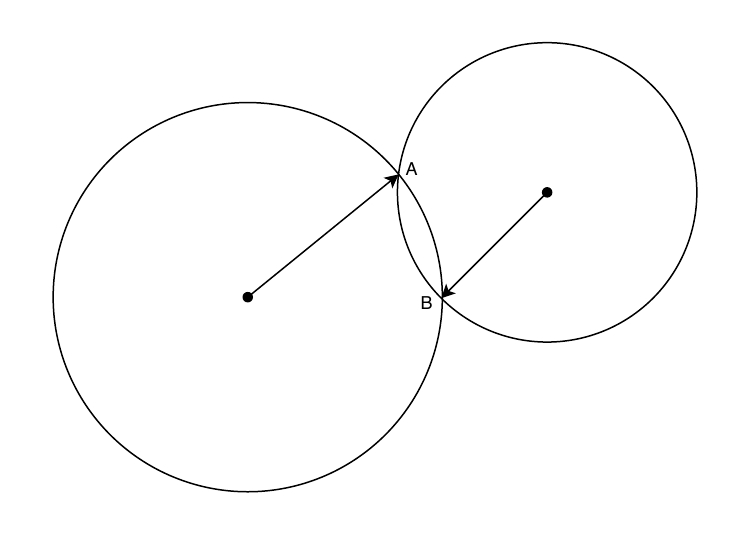}
  \caption{A schematic diagram solving the above system of
  equations, and also depicting the solutions in the parameter
  space after the Hough Transform.}\label{fig:ghough}

\end{figure}

The rotational axis of Earth undergoes precession in an approximately 26,000-year cycle, and the varying positions of the north celestial pole at different epochs are illustrated in Fig. \ref{fig:ghough}. Historical positions of the north celestial pole are calculated using the precession algorithm developed by Jan Vondrák~\cite{2011A&A...534A..22V,2012A&A...541C...1V}.

\subsection{Calculation Methods and Steps}

We assume that the observational records within each dataset are generated over a relatively short time period, implying that their \QJD values are grounded on the same zero point. These records are denoted by their coordinates in the J2000 epoch as $(\alpha_i ,\delta_i)$$(i = 1, 2,..., m)$, and can be converted into Cartesian co-ordinates as ($x_i ,y_i , z_i$ )($i = 1, 2,..., m$). The associated \QJD data are represented as $\omega_i$ ($i = 1, 2,..., m$).

Assuming the position of the north celestial pole during this time period is denoted as $(\alpha,\delta)$, it can be converted to Cartesian coordinates as $\vp$ $(x,y,z)$.

For any two stars, A and B, each characterized by spatial positions $\vs_i (x_i ,y_i , z_i)$ and $\vs_j (x_j , y_j ,z_j )$, where $i \neq j$, the relationship is expressed as

\begin{equation}
  \left\{
\begin{aligned} 
  \vs_i \cdot \vp &= \cos \omega_i \\
  \vs_j \cdot \vp &= \cos \omega_j \\
  \lvert\vp\rvert &= 1 
\end{aligned}
\right.
\end{equation}

which can be represented in the Cartesian coordinate system as

\begin{equation}
  \left\{
  \begin{aligned}
    x_i x + y_i y + z_i z &=  \cos \omega_i \\
    x_j x + y_j y + z_j z &=  \cos \omega_j \\
    x^2+y^2+z^2 &= 1 
  \end{aligned}
  \right.
  \end{equation}

To resolve the system of equations described above, as illustrated in Fig. \ref{fig:precession} in which the points on the blue line represent the positions of the north pole for different years, three potential scenarios for solutions exist: a unique solution, double solution, and no solution. Employing iterative methods facilitates the calculation of intersection solutions between any two stars.

Considering potential errors in observational data, it becomes imperative to introduce a certain degree of error sampling during the process of solving the system of equations. This practice proves effective in augmenting the precision of the results.

Owing to the precision issues in historical data observations, it is possible to introduce a certain amount of random error into the calculations for modeling purposes. By incorporating a large amount of random data within a certain margin of error, it is possible to enhance the results of data calculations. The result, observed in the phase space of right ascension and declination, corresponds to the maximum probability value near the final position of the north pole.

\subsection{Preprocessing of the Raw Data}

Prior to data processing and analysis, the declination data need to be converted. The degrees in the observational data are represented in the traditional ancient degree system (\textit{gudu} \kaiti{古度}, Chinese Degree), wherein a full circle corresponds to 365.25° in the celestial sphere. Consequently, as part of the data processing, a conversion is required to align the data with the modern system, where a full circle corresponds to 360°, using the relation

$$1~\gudu = \frac{360}{365.25} ~\textit{Babylonian Degrees}$$

\section{REFERENCE DATA}

During the data processing, we used the latest celestial measurement framework and modern stellar position data for our calculations. The calculation library employed in this article is the IAU's Standards of Fundamental Astronomy (SOFA)\citep{SOFA,gofa}, which guarantees accurate results.

The modern stellar catalog data used in our study are sourced from the Hipparcos Catalog~\citep{1997ESASP1200.....E}, the updated Hipparcos Catalog from 2007~\citep{2007A&A...474..653V}, and the extended Hipparcos Catalog (XHIP) from 2012~\citep{2012AstL...38..331A}. XHIP is derived from the Hipparcos Catalog but integrates the latest stellar proper motion, parallax, and radial velocity data. This comprehensive dataset allows for the accurate calculation of stellar positions across various historical epochs. The epoch for these catalogs is J2000.0~\citep{xinghan.100877}.

Once an approximate date for an ancient star catalog is established, further calculations incorporating proper motion and other relevant data are used to precisely determine the positions of stars within that catalog during the specific historical era. This additional step serves to validate and confirm the epoch of the data under consideration.

\section{DATA PROCESSING AND ANALYSIS}

We conduct an analysis on the two sets of star catalogs using the algorithm described above.

\subsection{Song Dynasty Star Catalog}

During the Northern Song Dynasty (960-1127 CE), there was a flourishing of society, the economy, and scientific and cultural pursuits. In the realm of astronomy and calendrics, astronomers actively engaged in the manufacture and calibration of astronomical instruments, undertaking extensive observations of celestial phenomena. Their remarkable achievements and records stand as a testament to the zenith of ancient Chinese technology and knowledge during that era.

Among the extant historical documents from the Song Dynasty, notable works such as \bookt{Lingtai Miyuan} \cbookt{灵台秘苑}, \bookt{Wenxian Tongkao} \cbookt{文献通考}, \bookt{Guankui Jiyao} \cbookt{管窥辑要}, \bookt{Qianxiang Tongjian} \cbookt{乾象通鉴}, and \bookt{Tianyuan Lili} \cbookt{天元历理} contain a wealth of stellar observation records. Pan Nai and others~\citep{hycatalog1,hycatalog2} meticulously organized and analyzed these documents, including a Ming Dynasty copy of \bookt{Lingtai Miyuan} \cbookt{灵台秘苑} housed in the National Library of China. Their efforts resulted in the compilation of a stellar catalog containing data for 360 stars, known as the \bookt{Song Huangyou Star Catalog} \cbookt{宋皇佑星表}. In this section, this catalog serves as the foundation, and the Hough transform algorithm is applied for the purpose of reprocessing it to ascertain its epoch.

\begin{table*}[ht]
  \centering
  \footnotesize
  \caption{Results of the Song Dynasty star catalog processing}
  \label{tab:song}
  \tabcolsep 5pt 
  \begin{tabular*}{\textwidth}{crcrcrcrcrcrcrcrc}
  \toprule
   Data Source & \multicolumn{2}{c}{verified value} & \multicolumn{2}{c}{Lingtai Miyuan 1} & \multicolumn{2}{c}{Lingtai Miyuan 2} & \multicolumn{2}{c}{Wenxian Tongkao} & \multicolumn{2}{c}{Xianglin} &  \multicolumn{2}{c}{Guankui Jiyao} & \multicolumn{2}{c}{Tianyuan Lili }\\ \hline
   Stars & \multicolumn{2}{c}{352} & \multicolumn{2}{c}{342} & \multicolumn{2}{c}{347} & \multicolumn{2}{c}{263} & \multicolumn{2}{c}{297} & \multicolumn{2}{c}{340} & \multicolumn{2}{c}{300} \\ \midrule
   \QJD~error & Epoch & $P_{dist}$ & Epoch & $P_{dist}$& Epoch & $P_{dist}$& Epoch & $P_{dist}$& Epoch & $P_{dist}$& Epoch & $P_{dist}$& Epoch & $P_{dist}$ \\ 
   (\gudu) & & ($deg$) & & ($deg$) & & ($deg$)& & ($deg$)& & ($deg$)& & ($deg$)& & ($deg$) \\ \hline
   0.10 &   995 & 0.03 & 989  & 0.11 & 987  & 0.07 & 1003 & 0.07 &  998 & 0.07 &  983 & 0.11 & 1002 & 0.08 \\
   0.20 &  1005 & 0.04 & 996  & 0.01 & 996  & 0.06 & 1015 & 0.06 & 1006 & 0.05 &  997 & 0.15 & 1012 & 0.07 \\
   0.25 &  1011 & 0.04 & 1002 & 0.00 & 1002 & 0.06 & 1022 & 0.05 & 1012 & 0.04 & 1009 & 0.16 & 1019 & 0.06 \\
   0.30 &  1016 & 0.05 & 1007 & 0.01 & 1008 & 0.07 & 1028 & 0.04 & 1018 & 0.03 & 1019 & 0.17 & 1026 & 0.05 \\
   0.35 &  1019 & 0.05 & 1010 & 0.02 & 1012 & 0.07 & 1033 & 0.04 & 1025 & 0.03 & 1026 & 0.18 & 1033 & 0.04 \\
   0.40 &  1021 & 0.05 & 1013 & 0.03 & 1016 & 0.07 & 1034 & 0.03 & 1028 & 0.03 & 1029 & 0.17 & 1036 & 0.04 \\
   0.45 &  1022 & 0.06 & 1013 & 0.03 & 1019 & 0.08 & 1033 & 0.02 & 1032 & 0.03 & 1030 & 0.16 & 1036 & 0.03 \\
   0.50 &  1022 & 0.06 & 1014 & 0.04 & 1020 & 0.08 & 1035 & 0.02 & 1031 & 0.03 & 1030 & 0.16 & 1036 & 0.03 \\
  \bottomrule
  \end{tabular*}
\end{table*}

\begin{figure}[htbp]
  \centering
  \includegraphics[width=\columnwidth]{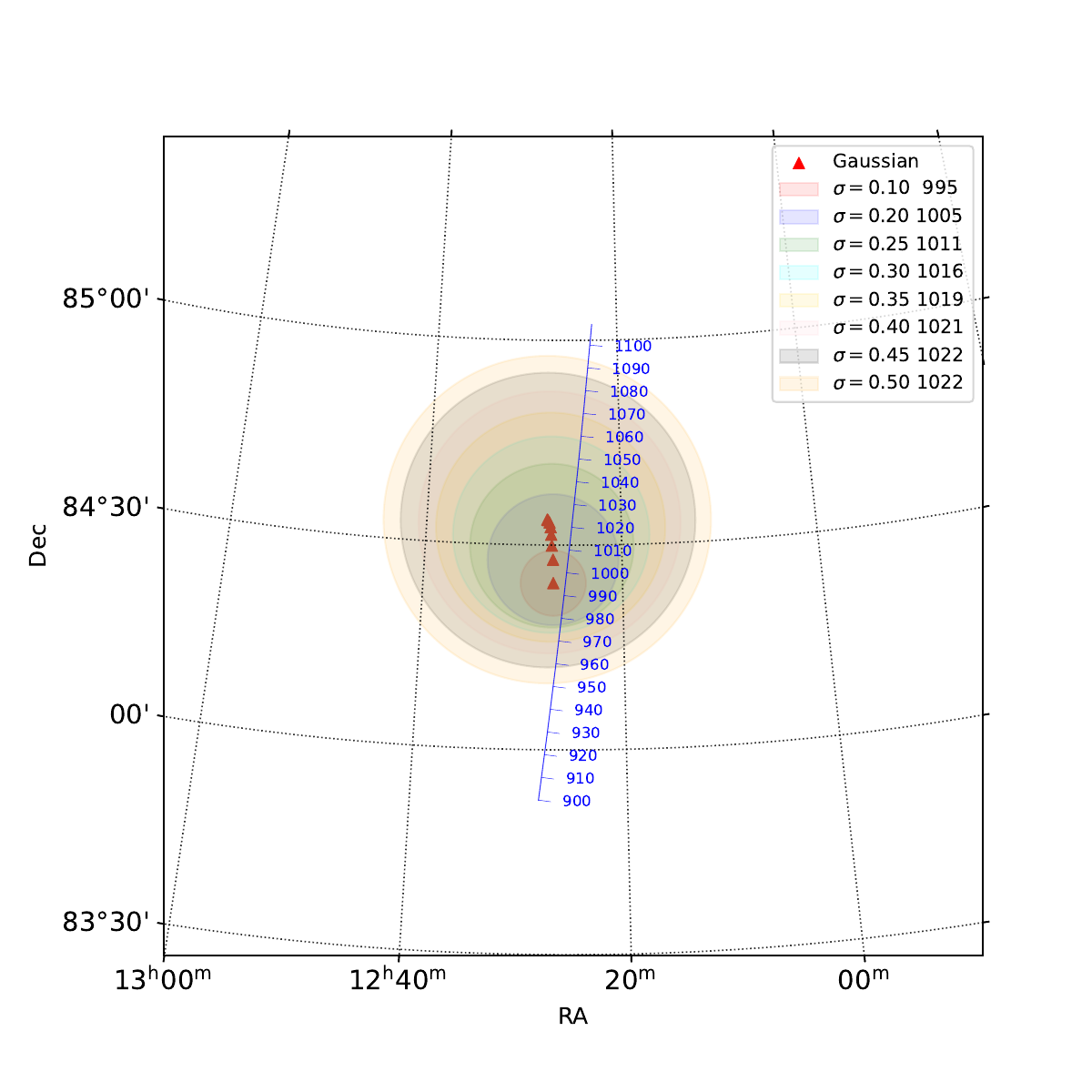}
  \caption{Results shown by colored circles, with radii
  corresponding to calculated error. Positions of the north pole
  at different epochs are shown in blue.}
  \label{fig:song-1011s}

\end{figure}

\begin{figure}[htbp]
  
  \centering
  \includegraphics[width=\columnwidth]{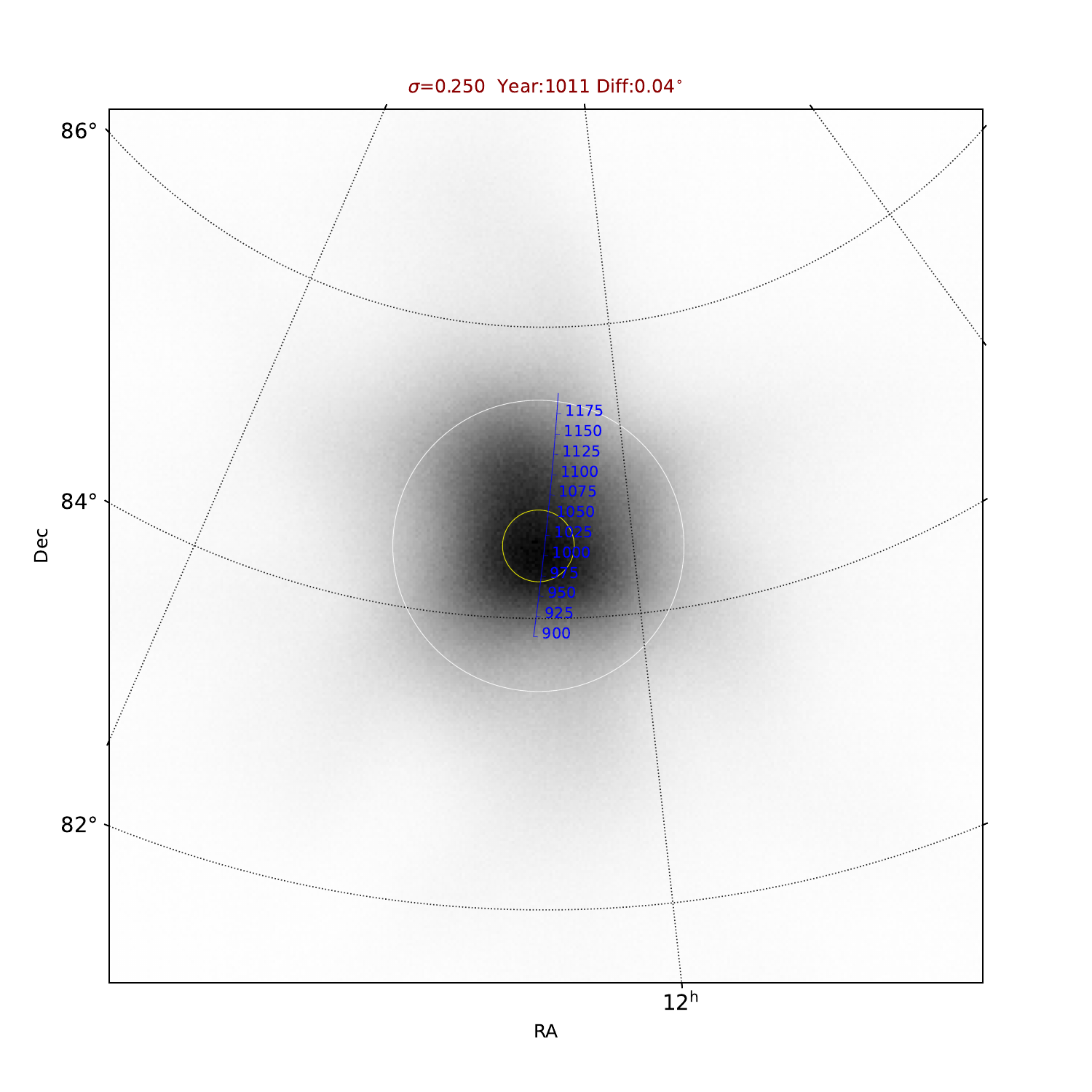}
  \caption{The optimal candidate time point is 1 011 CE. $\sigma=0.25$(\gudu)}
  \label{fig:song-1011}

\end{figure}

The calculation results are presented in Table \ref{tab:song}, where $P_{dist}$ denotes the distance from the candidate point to the nearest north pole. Fig. \ref{fig:song-1011s} illustrates that selecting different values for $\sigma$ yields distinct yet approximately similar candidate outcomes. Notably, in the cases of $\sigma = 0.25$ gudu and $\sigma = 0.3$ gudu, the convergence of candidate points appears relatively favorable. Fig. \ref{fig:song-1011} specifically represents the scenario with $\sigma = 0.25$ \textsf{gudu}. Consequently, it is reasonable to infer that the epoch falls approximately between 1011 CE and 1016 CE.

In the year 1011 CE, during the fourth year of Emperor Zhenzong's Dazhong Xiangfu \kaiti{大中祥符} reign in the Song Dynasty, as documented in \bookt{Song Huiyao Jigao} \cbookt{宋会要辑稿}~\citep{Songhyjg}, historical records note, In the third year of Dazhong Xiangfu, on the leap second of the second month, the Bureau of Astronomy reported, 'Winter official Han Xianfu \kaiti{韩显符} completed the casting of the copper celestial globe.' The decree mandated its transfer to the Longtu Pavilion and instructed Xianfu to mentor selected students in the craft. The same historical source further details, In the third year of Xiangfu, on the fifth day of the seventh month, Winter official Han Xianfu of the Bureau of Astronomy reported the positions of the outer stars relative to the polar degrees. The results were compiled into the volume 'Chongwen Mu Xiu Yao Du Fen Yu Ming Lu.' This information is also referenced under the entry 'Xiangfu Stellar Degrees' in the \bookt{Yuhai} \cbookt{玉海}\citep{yuhai}.

According to historical records, \textit{Han Xianfu} played a significant role in the production of the Copper Celestial Globe at the Zhidao Bureau in the first year of Zhidao (995). Engaged in crafting this celestial globe, he meticulously documented production methods, as recorded in the \textit{Song Shi · Tianwen Zhi}~\citep{tws22}, The Copper Seasonal Instrument was crafted by \textit{Han Xianfu} based on the methods passed down from the Tang Dynasty astronomer \textit{Li Chunfeng} \kaiti{李淳风} and the monk \textit{Yixing} \kaiti{僧一行}. Subsequently, he compiled these details into a ten-volume book and preserved it in the imperial library. \textit{Han Xianfu} passed away in the sixth year of Dazhong Xiangfu, and his student, the Song Dynasty astronomer \textit{Yang Weide}, was entrusted with compiling the \bookt{Jingyou Qianxiang Xinshu} \cbookt{乾象新书} (New Book of Jingyou and Qianxiang). During the writing of this book, \textit{Yang Weide} and his colleagues conducted a series of celestial observations.

It can be inferred that in approximately 1010 CE, the Winter official \textit{Han Xianfu} supervised the creation of the Xiangfu Longtu Pavilion Copper Celestial Globe, following Emperor Zhenzong's instruction to teach and leave behind a volume of observational records. His student, \textit{Yang Weide}, carried on \textit{Han Xianfu}'s legacy and organized additional stellar observations. The data recorded in extant historical documents likely represent records of these observations, partially transmitted through subsequent observations over several decades.

\subsection{Yuan Dynasty star catalog}

The National Library of China houses a Ming Dynasty (1368-1644 CE) copy of \bookt{Tianwen Huichao} \cbookt{天文汇抄} (Collection of Astronomical Material), comprising eleven sections in six volumes~\citep{twhc}. The third volume, titled \bookt{Sanyuan lieshe ruxiu quji ji} \cbookt{三垣列舍入宿去极集} (Collection of Coordinates of Stars in the Three Yuan and Twenty-Eight Xiu), contains detailed observational records of approximately 740 stars.

\begin{table}[ht]
  \centering
  \footnotesize
  \caption{Yuan Dynasty star catalog (Tianwen Huicaho catalog) processing results}
  \label{tab:gsj}
  \tabcolsep 10pt 
  \begin{tabular*}{0.9\columnwidth}{ccccc}
  \toprule
   \QJD~error(\gudu) & Epoch & $P_{dist}$($deg$) \\\hline
   0.050 & 1353 & 0.20 \\ 
   0.100 & 1355 & 0.20 \\ 
   0.125 & 1357 & 0.20 \\ 
   0.200 & 1361 & 0.20 \\ 
   0.250 & 1363 & 0.20 \\ 
   0.333 & 1366 & 0.20 \\
  \bottomrule 
  \end{tabular*}
\end{table}

After its discovery and organization in the 1980s, several scholars undertook research on this text. In 1986, \textit{Chen Ying} analyzed the data and, as a result of \textit{Guo Shoujing}'s \kaiti{郭守敬} stellar observations, concluded that the star catalog's observations were made in the Yuan Dynasty (1279-1368 CE) in 1280 CE~\citep{chenyingtwhc}. In the 1990s, \textit{Sun Xiaochun} conducted meticulous data analysis, determining the actual observation year to be approximately 1380 CE, during the early Ming Dynasty's Hongwu period~\citep{sunxc1996twhc}. He published the identification results of 678 stars. \textit{Pan Nai} also studied and analyzed the data, reaching a conclusion similar to \textit{Chen}'s, dating it to approximately 1280 CE~\citep{PanNai2009HCSO}.

\begin{figure}[ht]
  \centering
  \includegraphics[width=\columnwidth]{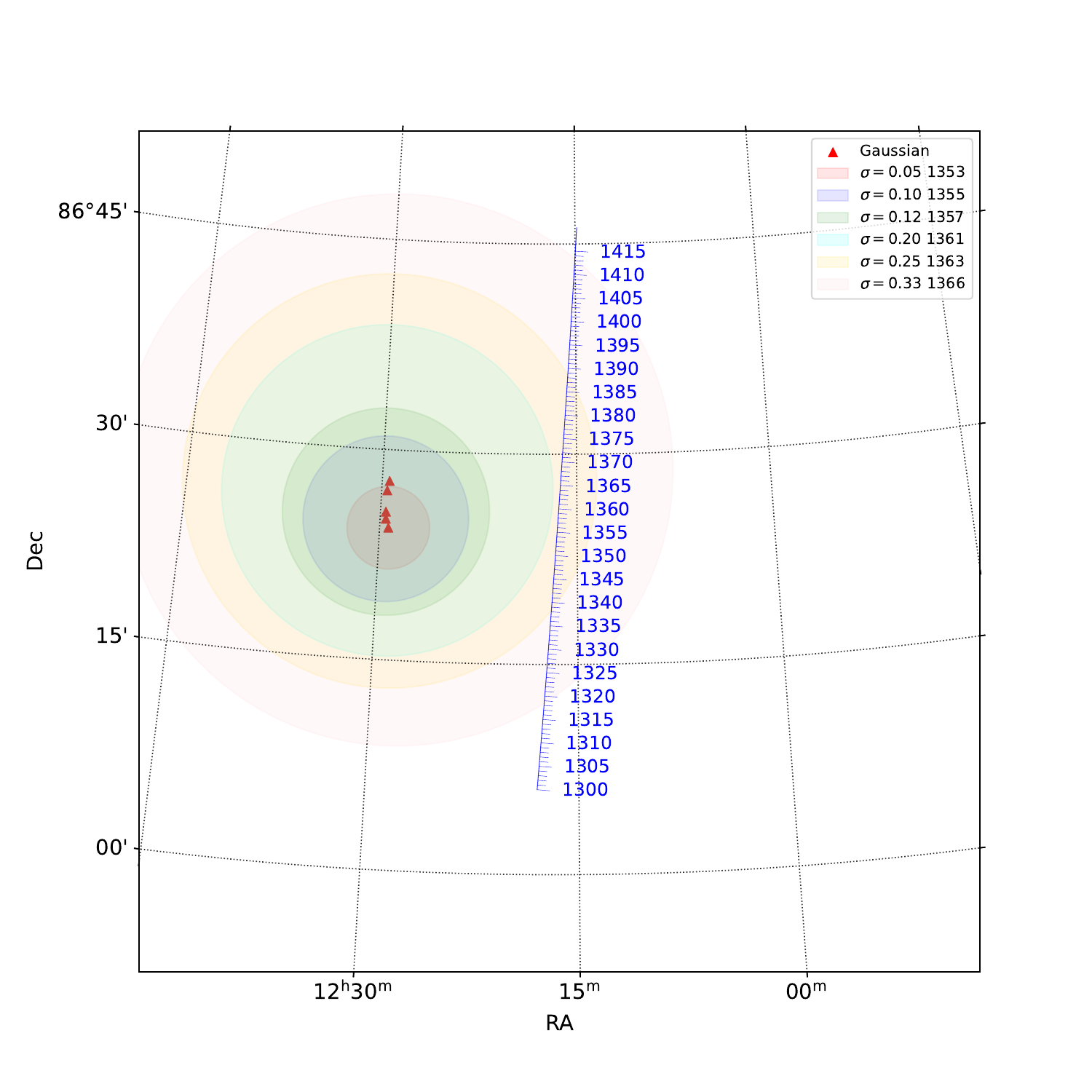}
  \caption{Results, calculated with different error radii, are
  represented by colored circles. The positions of the north pole
  at different epochs are shown in blue.}
  \label{fig:gsj-1355s}

\end{figure}
\begin{figure}[ht]
  \centering
  \includegraphics[width=0.9\columnwidth]{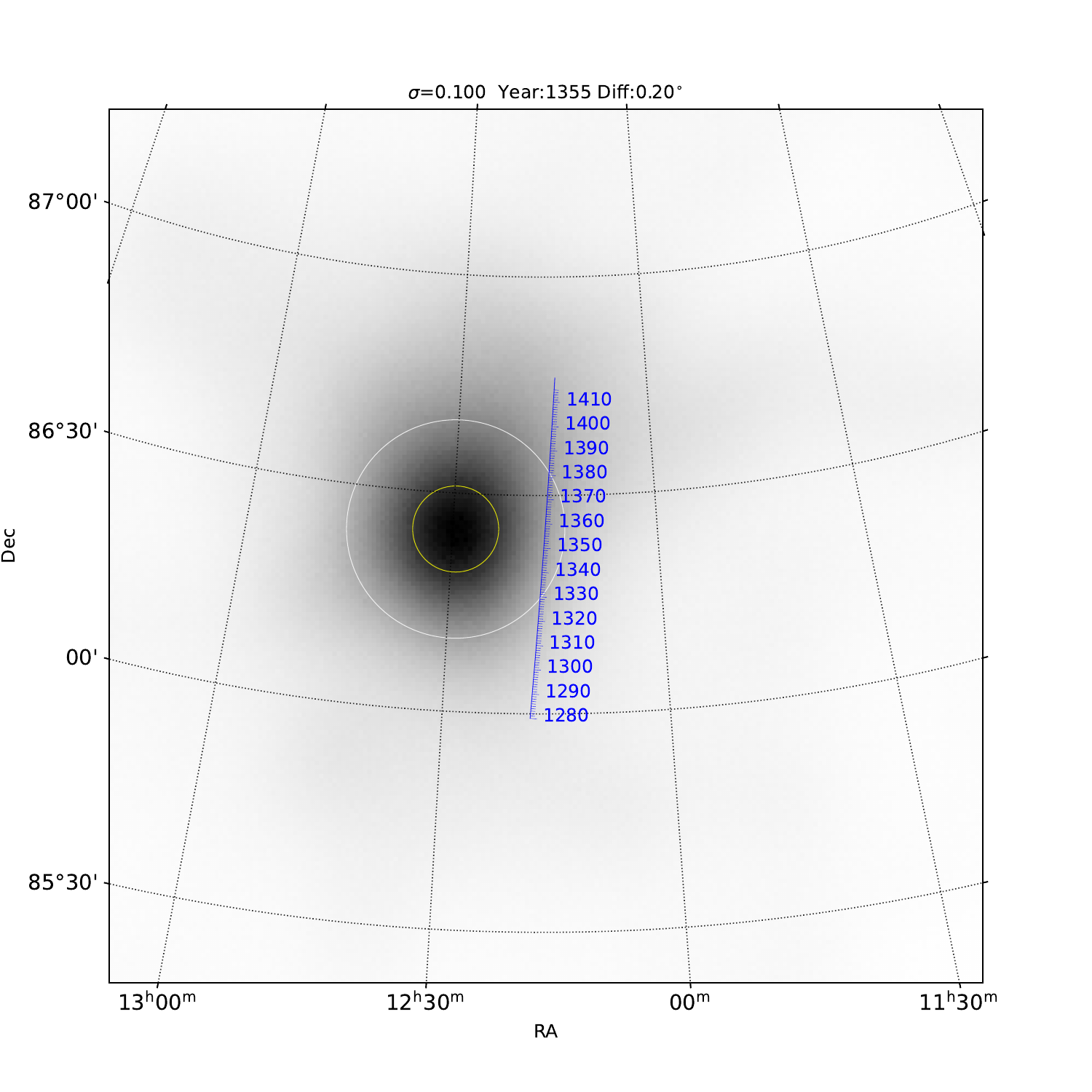}
  \caption{The optimal candidate year point is 1355 CE.}
  \label{fig:gsj-1355}
\end{figure}

In this section, calculations use the data of the 669 stars published by \textit{Sun Xiaochun}. Error ranges for declination are sampled using a Gaussian distribution. Table \ref{tab:gsj} presents the calculated years and their different error radii, along with the corresponding distances ($P_{dist}$) between the calculated and actual north pole positions for each year. Fig. \ref{fig:gsj-1355s} provides visual representations at various error radii, while Fig. \ref{fig:gsj-1355} shows computational results for the optimal case with $\sigma = 0.1$ (\gudu). Consequently, it can be reasonably inferred that the epoch is approximately 1355 CE.

The year 1355 CE corresponds to the fifteenth year of the Yuan Dynasty's Zhizheng \songti{至正} era, marking the latter part of the Yuan Dynasty. Judging from the data records, this set is exceptionally rare. The data is presented in the form of star charts, with the positional data of stars meticulously recorded alongside them. Furthermore, the data are organized into a percentage-based system, which was uncommon for that era. It can be speculated that this represents a meticulous stellar observation conducted by officials of the Bureau of Astronomy toward the end of the Yuan Dynasty, utilizing the celestial globe created by \textit{Guo Shoujing} \kaiti{郭守敬}. These observations were preserved in chart form and subsequently copied during the Ming Dynasty.

\section{CONCLUSIONS}

Based on the calculations and analysis of the two sets of data above, an initial evaluation of the Generalized Hough Transform algorithm for determining epochs can be provided. This method proves effective in estimating the observation epochs of ancient Chinese astronomical records and demonstrates greater accuracy compared with previous research efforts. The advantages of using the Hough transform are its independence from right ascension data and its ability to handle erroneous data without specific adjustments. Through extensive error sampling, the impact of incorrect data can be mitigated. Simultaneously, in the calculation process, various parameters related to stellar proper motion are taken into account, further enhancing the accuracy of epoch determination.

In the calculation process of this study, observational data from the Song Dynasty show better data clustering under a condition of $\sigma = 0.25$ gudu, while in the Yuan Dynasty, under a condition of $\sigma = 0.1$ gudu, the data clustering surpasses that of the Song Dynasty. This suggests that from the Song Dynasty to the Yuan Dynasty, with the advancement of instruments, the precision of the data has significantly improved.

The method employed in this study has yielded favorable results on the Song-Yuan Star Catalog and offers a new perspective for analyzing ancient astronomical catalog data in the future.

\bibliographystyle{gbt7714-numerical}
\bibliography{refs.bib}

\end{document}